\documentclass[aps, prd, amsmath, floats, floatfix, twocolumn, nofootinbib, superscriptaddress]{revtex4-1}
\usepackage[pdftex]{graphicx}
\usepackage{color}
\usepackage{latexsym}
\usepackage[colorlinks]{hyperref}
\usepackage{amsthm,amsmath,amssymb}
\usepackage{mathrsfs}

\newcommand{\be}{\begin{eqnarray}}
\newcommand{\ee}{\end{eqnarray}}

\begin{document}

\title{Improved Constraints on Non-Kerr Deviations from Binary Black Hole Inspirals Using GWTC-4 Data}

\author{Debtroy~Das}
\email{ddebtroy22@m.fudan.edu.cn}
\affiliation{Center for Astronomy and Astrophysics, Department of Physics, Fudan University, Shanghai 200438, China}

\author{Swarnim~Shashank}
\email{swarnim.shashank@astro.uni-tuebingen.de}
\affiliation{Institut f\"ur Astronomie und Astrophysik, Eberhard-Karls Universit\"at T\"ubingen, D-72076 T\"ubingen, Germany}
\affiliation{College of Fellows, Eberhard-Karls Universit\"at T\"ubingen, D-72070 T\"ubingen, Germany}

\author{Cosimo~Bambi}
\email[Corresponding author: ]{bambi@fudan.edu.cn}
\affiliation{Center for Astronomy and Astrophysics, Department of Physics, Fudan University, Shanghai 200438, China}
\affiliation{School of Natural Sciences and Humanities, New Uzbekistan University, Tashkent 100000, Uzbekistan}

\begin{abstract}
Gravitational wave (GW) observations of binary black hole (BBH) mergers provide a unique opportunity to probe the nature of spacetime in the strong-field and dynamical regime. We present updated constraints on deviations from the Kerr metric using BBH inspirals from the fourth Gravitational-Wave Transient Catalog (GWTC-4). Building on our previous GWTC-3 analysis, we employ a theory-agnostic framework in which non-Kerr effects of the Johannsen metric are incorporated as parametrized corrections to the GW phase within the post-Newtonian framework. We perform Bayesian parameter estimation on a selected subset of GWTC-4 events to constrain the deformation parameters $\alpha_{13}$ and $\epsilon_3$, yielding significantly tighter bounds compared to earlier results. When varied individually, the deformation parameters are found to be consistent with zero, providing no evidence for departures from the Kerr geometry. Our results reinforce the validity of General Relativity, particularly the Kerr hypothesis, and highlight the progress enabled by GWTC-4.
\end{abstract}

\maketitle

%%%%%%%%%%%%%%%%% BODY OF PAPER %%%%%%%%%%%%%%%%%%

\section{Introduction}

The direct detection of gravitational waves (GWs) by the LIGO–Virgo–KAGRA (LVK) collaboration~\cite{abbott2016gw150914,abbott2016gw150914a,LIGOScientific:2016aoc} has opened a unique window into the dynamics of compact and energetic systems, particularly in regions and environments inaccessible to electromagnetic probes. Binary black hole (BBH) mergers, in particular, provide a pristine laboratory to test the predictions of General Relativity (GR) in highly dynamical and nonlinear regimes~\cite{LIGOScientific:2019fpa,LIGOScientific:2020tif,LIGOScientific:2021sio,LIGOScientific:2026qni,LIGOScientific:2026fcf,LIGOScientific:2026wpt,krishnendu2021testing,LIGOScientific:2016lio,li2012towards,Maggio:2022hre,Ghosh:2017gfp,Carson:2019kkh,Agathos:2013upa,Krishnendu:2017shb,Arun:2006yw,Das:2025riq}. Within GR, astrophysical black holes are expected to be described by the Kerr metric, fully characterized by their mass and spin. Any statistically significant deviation from this description would indicate physics beyond GR or a breakdown of the Kerr hypothesis.

GR tests can be broadly categorized into two wide classes: theory-specific and theory-agnostic approaches. In the former, one compares GR directly with alternative theories, while in the latter, one looks for generic deviations from GR predictions without committing to a specific theory or model~\cite{bambi2017black}. Parametrized frameworks~\cite{LIGOScientific:2026fcf,Maggio:2022hre,Carson:2019kkh,konoplya2016general,ghasemi2016note}, such as the parametrized post-Einsteinian (ppE) formalism~\cite{shashank2022constraining,cardenas2020gravitational,Yunes_Yagi_Pretorius_2016,Yunes_Pretorius_2009,nissanke2005gravitational}, provide a powerful tool to search for such deviations in a model-independent way.

In our previous work~\cite{Das:2024mjq}, we adopted a theory-agnostic approach to test the Kerr hypothesis by introducing deviations in spacetime, considering the Johannsen metric~\cite{johannsen2013regular}. By incorporating leading-order corrections to the GW phase up to 4 post-Newtonian (PN) order, we constrained the deformation parameters using BBH events from the third Gravitational-Wave Transient Catalog (GWTC-3)~\cite{ligo2021gwtc}. Our results showed that, while individual deformation parameters can be constrained when varied independently, strong degeneracies prevent simultaneous constraints.

Since then, the LVK collaboration has released the fourth Gravitational-Wave Transient Catalog (GWTC-4)~\cite{LIGOScientific:2025slb}, which includes 128 compact binary merger events with improved detector sensitivity and relatively higher signal-to-noise ratios (SNRs). This provides an opportunity to revisit and refine previous constraints on deviations from the Kerr metric.

While the GWTC-4 dataset significantly broadens the scope of discovered BBH properties, our study focuses on a population of binaries with comparably similar properties to those analyzed in the previous study using the GWTC-3 catalog. Because the LVK detectors have become more sensitive, they can detect a greater number of events with higher SNRs, leading to a considerably larger sample set of GW events than the one used in the previous study.

In this work, we perform a follow-up analysis using BBH inspiral data from GWTC-4. The mathematical framework and waveform modeling remain identical to our earlier study. We summarize the main idea in Section~\ref{sec2}. The key difference lies in the expanded dataset, which enables tighter constraints on the deformation parameters and improved statistical robustness. Our goal is to assess how the increased data volume and quality impact the bounds on non-Kerr deviations and to evaluate the consistency of GW observations with the Kerr hypothesis. The data analysis and results are presented in Section~\ref{sec3}, followed by concluding remarks in Scetion~\ref{sec4}

\section{Mathematical Framework}\label{sec2}

\subsection{Parametrized deviations from the Kerr metric and Waveform modifications}

We model potential deviations from the Kerr spacetime using the Johannsen metric~\cite{johannsen2013regular}. This metric is a phenomenological extension that introduces deformation parameters while preserving key properties such as stationarity, axisymmetry, asymptotic flatness, and particle motion still possesses a Carter-like constant. In the limit where all deformation parameters vanish, the metric reduces to the Kerr solution.

Following our previous work~\cite{Das:2024mjq}, we focus on the deformation parameters $\alpha_{13}$ and $\epsilon_3$. The presence of these non-Kerr deviations modifies the dynamics of BBH inspirals. Consequently, the phase evolution of the emitted GWs is also modified. We compute these corrections within the parametrized ppE framework, where deviations enter as additional terms in the GW phase~\cite{shashank2022constraining,cardenas2020gravitational,nissanke2005gravitational}. We restrict our analysis to the quasi-circular inspiral approximation. The total GW phase can be written schematically as a sum of the GR contribution and a non-Kerr correction term
\be \label{eq:phase_gr_nk}
\Psi_{\mathrm{GW}}(f)=\Psi_{\mathrm{GW}}^{\mathrm{GR}}(f) + \Psi_{\mathrm{GW}}^{\rm NK}(f) \, .
\ee
The non-Kerr contribution, labeled by $\rm NK$ superscript, is derived by modifying the orbital dynamics through corrections to the binding energy and orbital frequency, obtained from the deformed metric. Assuming that dissipative effects remain unchanged from GR at leading order, the corrections enter primarily through the conservative sector. As in our earlier study~\cite{Das:2024mjq}, we retain terms up to 4PN order in the phase expansion following Ref.~\cite{Carson:2020iik}. This ensures consistency with the waveform approximant used and allows us to capture the dominant effects of the deformation parameters on the inspiral signal. The additional terms in GW phase due to non-vanishing $\alpha_{13}$ and $\epsilon_3$ are 
\be \label{eq:phase}
\Psi_{\mathrm{GW}}^{\rm NK} &=& 
-\frac{45 \alpha_{13} \kappa \sqrt[3]{u}}{4 \eta^{6/5}}
+\frac{10 \alpha_{13} \kappa u}{3 \eta ^{8/5}}
-\frac{15 \alpha_{13} \kappa }{4 \eta ^{4/5}\sqrt[3]{u}} \nonumber\\
&& -\frac{10 \alpha_{13} \kappa u \log (u)}{3 \eta ^{8/5}} \, . \\
\Psi_{\mathrm{GW}}^{\rm NK} &=& 
\frac{45 \epsilon_{3} \kappa \sqrt[3]{u}}{8 \eta ^{6/5}}
-\frac{5 \epsilon_{3}\kappa u}{3 \eta ^{8/5}} 
+\frac{15 \epsilon_{3} \kappa }{32 \eta ^{4/5} \sqrt[3]{u}} \nonumber\\
&& +\frac{5 \epsilon_{3} \kappa u \log (u)}{3 \eta ^{8/5}} \, . 
\ee
Here, $\eta = \mu/m$ is the symmetric mass ratio ($\mu$ is the reduced mass and $m$ is the total mass), $u = \eta^{3/5}\pi m f$, and $f$ is the Fourier frequency.

By inspecting the functional forms of the non-Kerr phase corrections, a strong degeneracy between $\alpha_{13}$ and $\epsilon_3$ can be anticipated. Both parameters map onto the exact same PN orders. The coefficients for the $u^{1/3}$, $u$, and $u \log u$ terms in $\Psi_{\mathrm{GW}}^{\rm NK}(\alpha_{13})$ are -2 times those in $\Psi_{\mathrm{GW}}^{\rm NK}(\epsilon_3)$. Since the phase terms have opposite signs in the two equations, the deformation parameters must have the same sign as each other to cancel out. Hence, we expect strong positive degeneracy between $\alpha_{13}$ and $\epsilon_3$ in the parameter estimation, which typically manifests as an elongated linear correlation structure in their joint posterior distributions. The only thing keeping this degeneracy from being 100\% perfect, which would make the parameters completely unmeasurable when varied together, is the $u^{-1/3}$ term, where the ratio is $-8$ instead of $-2$. This mismatch may allow parameter estimation routines to eventually separate the two signals given enough SNR. This trend has also been observed in Ref.~\cite{Das:2024mjq} (refer to Fig. 5 in Ref.~\cite{Das:2024mjq}).

The non-Kerr deviations considered modify only the inspiral portion of the signal. These inspiral corrections are stitched to a GR waveform. We incorporate these phase corrections into a frequency-domain waveform model based on the \textsc{IMRPhenom} family~\cite{Husa:2015iqa, Khan:2015jqa, Hannam:2013oca,bohe:2016ph}. As a baseline model, we adopt the \textsc{IMRPhenomXPHM}~\cite{Pratten:2020ceb} waveform model, ensuring compatibility with standard GW data analysis pipelines. During the analysis, only the inspiral portions of the strain data are analyzed.

\subsection{Bayesian inference framework}

We estimate the source parameters and deformation parameters using Bayesian inference~\cite{box2011bayesian,Veitch:2009hd,Romero-Shaw:2020owr}. Given GW data $d$ and model parameters $\theta$, the posterior distribution is given by
\be \label{eq:bayesian}
 p ( \theta \mid d) \propto \mathcal{L}(d \mid \theta) \pi_{\rm PE} (\theta).
\ee
The term $\pi_{\rm PE} (\theta)$ is the prior distributions and $\mathcal{L}(d \mid \theta)$ is the likelihood. The parameters $\theta$ include both the standard BBH parameters (masses, spins, distance, etc.) and the non-Kerr deformation parameters. $\mathcal{L}(d \mid \theta)$ determines the probability of GW data $d$ given a waveform model characterized by $\bm{\theta}$~\cite{Veitch:2009hd}. The likelihood function takes the form of
\be
\mathcal{L}(d \mid \bm{\theta}) \propto \exp \left(-\sum_{k} 
\frac{2 \left| d_{k}-h_{k}(\bm{\theta}) \right|^{2}}{T S_{k}} 
\right).
\ee
where $h(\bm{\theta})$ denotes the gravitational waveform for the BBH signal with parameters $\bm{\theta}$. $T$ is the duration of the data segment being analyzed and $S_{k}$ is the noise power spectral density (PSD) characterizing the sensitivity of the detector. $k$ signifies the frequency dependence of the data, the waveform, and the PSD.

\section{Data Analysis and Results}\label{sec3}

For this study, we consider BBH events from the GWTC-4 catalog. To ensure consistency with our previous analysis, we select only those events in which the inspiral part of the signal is well captured by the detectors. Hence, we apply selection criteria similar to those used in our earlier work~\cite{Das:2024mjq}. These include cuts on total mass in the detector frame ($M<80$~$M_\odot$), network SNR ($>10$), and spin-precession parameter ($\ensuremath{\chi_{\mathrm{p}}} < 0.5$)~\cite{LIGOScientific:2025slb}. Only events satisfying these criteria are included in the analysis. The final set consists of 11 BBHs from new events in GWTC-4.

These events primarily consist of nearly symmetric ($q \in [0.5, 1]$), low-spin (effective spins $\chi_{eff}$ tightly clustered around zero) binaries with chirp masses $\mathcal{M}$ ranging between $\sim 6-40 M_\odot$, with the majority concentrated in the $10-25 M_\odot$ range. As these events have low-to-moderate chirp masses, their inspiral phase is long and well covered within the sensitive band of the detectors. If we were analyzing heavier binaries (e.g., $ \mathcal{M} \sim 50- 100 M_\odot$), the detector would only observe the merger and ringdown parts, making an inspiral-only model completely invalid. The quasi-circular PN expansion behaves most reliably for comparable mass systems. Highly asymmetric systems or those with large precessing spins introduce complex modulations and higher-order harmonics that could cause heavy degeneracy with the ppE phase corrections. By focusing on symmetric, low-spin events, we minimize systematic modeling errors. Hence, we expect our bounds on the non-Kerr parameters to be robust.

We use real data strains made publicly available through the GW Open Science Center~\cite{LIGOScientific:2025snk} from BBH merger events observed by the LIGO–Hanford (H1) and LIGO–Livingston (L1) detectors~\cite{LIGOScientific:2014pky} (except for GW240109\_050431 and GW231231\_154016 which are observed by only LIGO–Hanford). The data strain used are obtained from channels: H1:GDS-CALIB\_STRAIN\_CLEAN\_AR and L1:GDS-CALIB\_STRAIN\_CLEAN\_AR. The attribute CLEAN in H1 and L1 indicates use of noise subtraction~\cite{Vajente:2019ycy,LIGO:2021ppb}. AR stands for Analysis Ready and indicates that these channels contain only data for times that are ready for analysis. Bayesian parameter estimation was performed using \textsc{Bilby}~\cite{Ashton:2018jfp} with the \textsc{Dynesty} nested-sampling algorithm~\cite{2020MNRAS.493.3132S}. We adopt prior distributions consistent with those used in the LVK analyses for BBH parameters~\cite{Ashton:2018jfp,Romero-Shaw:2020owr}, while using uniform priors for the deformation parameters within the interval $[-5,5]$, which lies in a physically motivated range. Furthermore, following the theoretical constraints derived in the metric paper~\cite{johannsen2013regular}, the physically motivated lower bounds for the prior ranges of $\alpha_{13}$ and $\epsilon_{3}$ are determined dynamically for each binary using Eq.~(19) of our previous work~\cite{Das:2024mjq}. These theoretical bounds are enforced via conversion functions in \textsc{Bilby} during the analysis.

\begin{figure}[h]
\centering
\includegraphics[width =\linewidth]{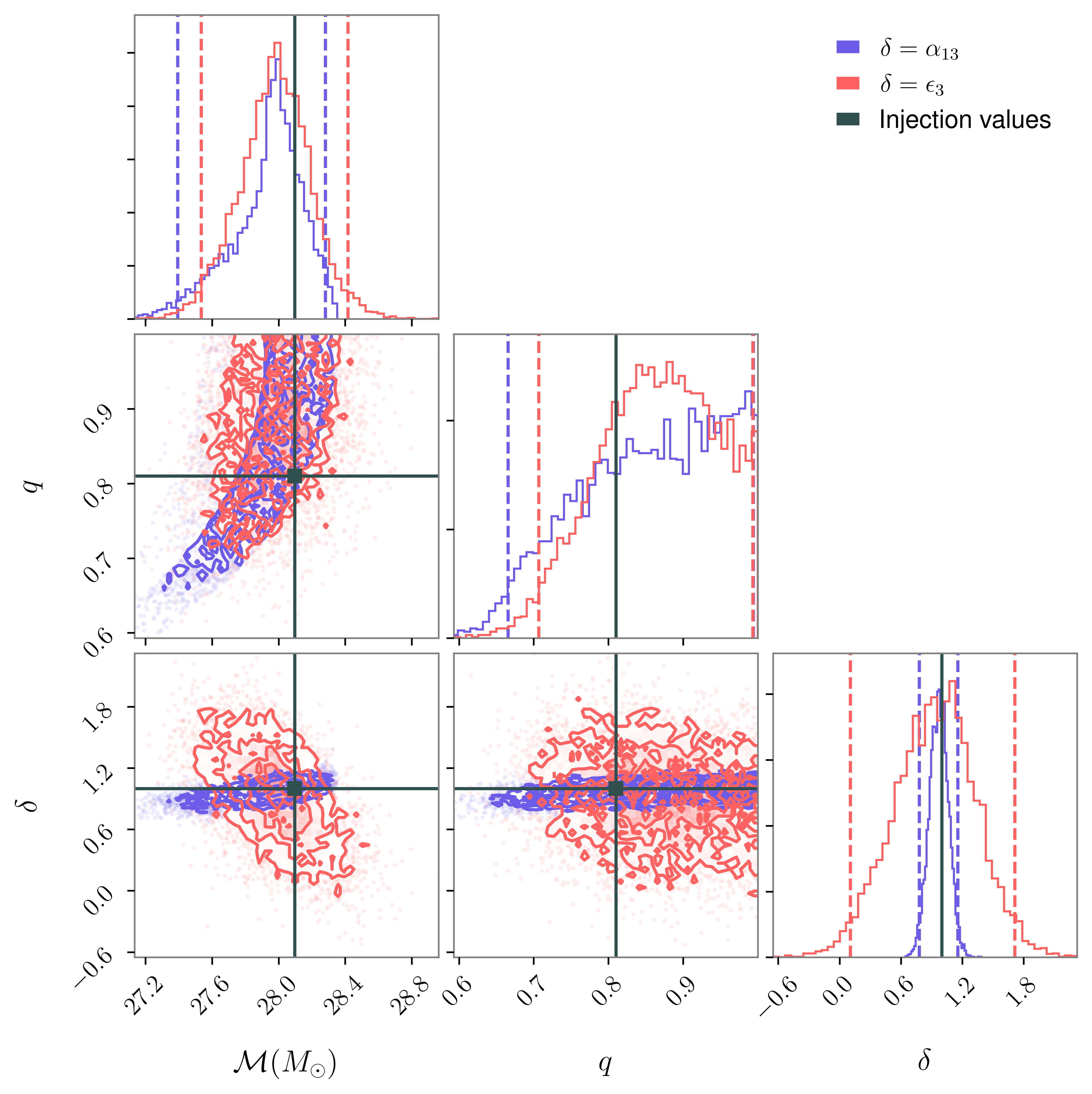} \\
\caption{Corner plots showing the chirp mass $\mathcal{M}$, mass ratio $q$ and the non-Kerr parameter $\alpha_{13}$ and $\epsilon_3$ for two GW150914-like \textsc{IMRPhenomXPHM} injections. The distributions shown in purple corresponds to the analysis when the injection assumes $\alpha_{13}=1$ and $\epsilon_3= 0$ and the distributions shown in pink corresponds to the analysis when the injection assumes $\epsilon_3=1$ and $\alpha_{13}= 0$. The solid slate colored lines represent the parameter values of the injection. The dashed lines represent the 2-$\sigma$ levels of the posterior distributions.} \label{fig:inj}
\end{figure}

To test whether the waveform template being used can recover synthetic non-Kerr waveforms, we conduct an injection study with a GW150914-like BBH system. We ran parameter estimation for 16 parameters in 2 cases. Two distinct injection scenarios are analyzed using the \textsc{IMRPhenomXPHM} waveform model. The first scenario is an injection assuming a non-zero deformation parameter $\alpha_{13}=1$ and $\epsilon_3=0$, while the second configuration assumes $\epsilon_3=1$ and $\alpha_{13}=0$. Figure~\ref{fig:inj} illustrates the posterior distributions for the intrinsic parameters, chirp mass $\mathcal{M}$ and mass ratio $q$, along with the parameterized non-Kerr deviations, $\alpha_{13}$ and $\epsilon_3$, which are visually unified under the label $\delta$. The posteriors from the first case are plotted in purple and the second case in pink. As evident from the corner plot, the parameter estimation framework successfully recovers the injected parameters well within the $2\text{-}\sigma$ credible intervals, indicated by the dashed lines. The true values of the injected signals, marked by the solid slate-colored lines ($\mathcal{M} = 28.096\,M_\odot$, $q=0.81$, and $\delta=1$), lie comfortably within the high-density regions of the respective posterior distributions. This excellent agreement between the injected values and the recovered posterior distributions highlights the robustness of our analysis.

For each selected event, we perform two independent analyses varying one deformation parameter while fixing the other to zero. We repeat the procedure for the second parameter. In all cases where a single deformation parameter is allowed to vary, the posterior distributions remain consistent with the GR prediction (i.e., zero deformation) within the statistical uncertainties. This reinforces the conclusion that current GW observations do not provide evidence for deviations from the Kerr metric. The results from the GWTC-4 analysis show a clear improvement over the constraints obtained from GWTC-3 presented in Ref.~\cite{Das:2024mjq}. Tighter posterior distributions for the deformation parameters are obtained from analysis of events with higher SNRs. 

Allowing both parameters to vary simultaneously results in neither being meaningfully constrained, as we observed in our previous study, so we do not display them in the paper. We again observe strong degeneracies between them which we anticipated by glancing at the functional forms of the non-Kerr phase terms. Given that the degeneracy is largely fundamental to their mathematical definition, it is extremely challenging to fully break it. This prevents meaningful independent constraints on both parameters at the same time. The correlation structure remains largely unchanged across all the events examined in our study. Due to the mismatch in the coefficient ratio of the $u^{-1/3}$ term, a GW event with exceptionally high SNR might be a suitable candidate to break the degeneracy.

The strong near perfect degeneracy between the non-Kerr parameters is strictly internal to the model.
Mild correlations of the non-Kerr parameters with chirp mass $\mathcal{M}$ exist. $\mathcal{M}$ exhibits positive correlation with $\alpha_{13}$ and negative correlation with $\epsilon_3$. This is primarily 
because the overall phase evolution is inherently coupled to the black hole masses. However, this does not prevent parameter convergence.

\begin{table}[t]
\centering
\renewcommand\arraystretch{1.5}
\begin{tabular}{lcc}
\hline\hline
Event & \hspace{1.0cm} $\alpha_{13}$ \hspace{1.0cm} & \hspace{1.0cm} $\epsilon_3$ \hspace{1.0cm} \\
\hline
GW230605\_065343 & $ -0.299 _{- 0.264 }^{+ 0.32 }$& $ 1.254 _{- 1.487 }^{+ 1.144 }$\\
GW230627\_015337 & $ 0.067 _{- 0.164 }^{+ 0.164 }$& $ -0.279 _{- 0.69 }^{+ 0.74 }$\\
GW230628\_231200 & $ 0.231 _{- 0.595 }^{+ 0.59 }$& \\
GW230814\_230901 & $ -0.029 _{- 0.353 }^{+ 0.332 }$& $ 0.124 _{- 1.457 }^{+ 1.411 }$\\
GW230904\_051013 & $ 0.784 _{- 1.117 }^{+ 0.954 }$& \\
GW230927\_153832 & $ -0.083 _{- 0.541 }^{+ 0.548 }$& $ 0.355 _{- 2.289 }^{+ 2.137 }$\\
GW231020\_142947 & $ 0.049 _{- 0.648 }^{+ 0.964 }$& $ -0.369 _{- 3.887 }^{+ 2.818 }$\\
GW231108\_125142 & $ 0.394 _{- 1.104 }^{+ 1.132 }$& \\
GW231226\_101520 & $ -0.091 _{- 0.704 }^{+ 0.757 }$& $ 0.357 _{- 3.044 }^{+ 2.965 }$\\
GW231231\_154016 & $ -0.289 _{- 0.639 }^{+ 0.802 }$& $ 1.212 _{- 3.11 }^{+ 2.729 }$\\
GW240109\_050431 & $ -0.213 _{- 1.164 }^{+ 1.621 }$& \\
\hline\hline
\end{tabular}
\caption{Summary of the non-Kerr parameters showing the medians and the 99\% intervals from our analysis.} 
\label{tab:99} 
\end{table}

\begin{figure*}[t]
\centering
\includegraphics[width =0.9\linewidth]{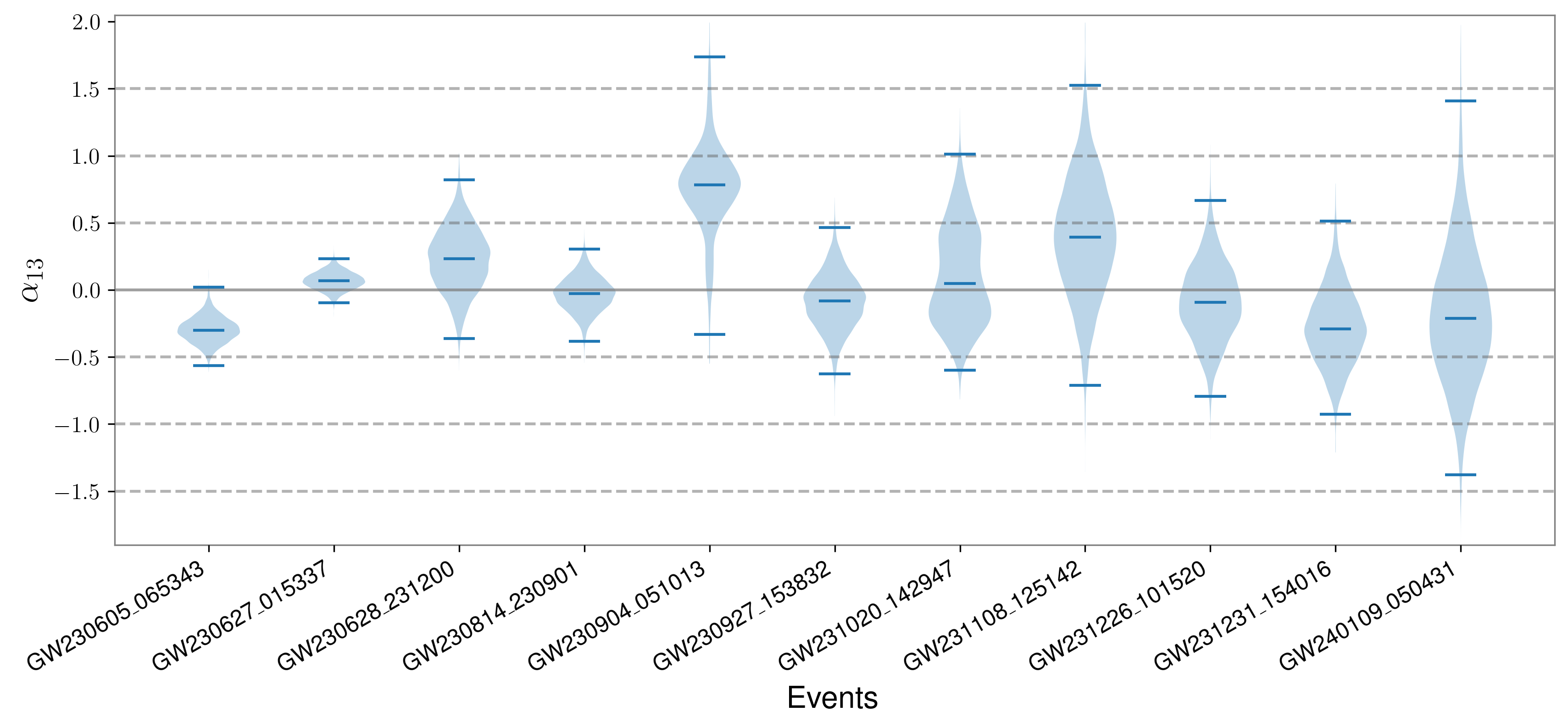} \\
\vspace{1.0cm}
\includegraphics[width =0.9\linewidth]{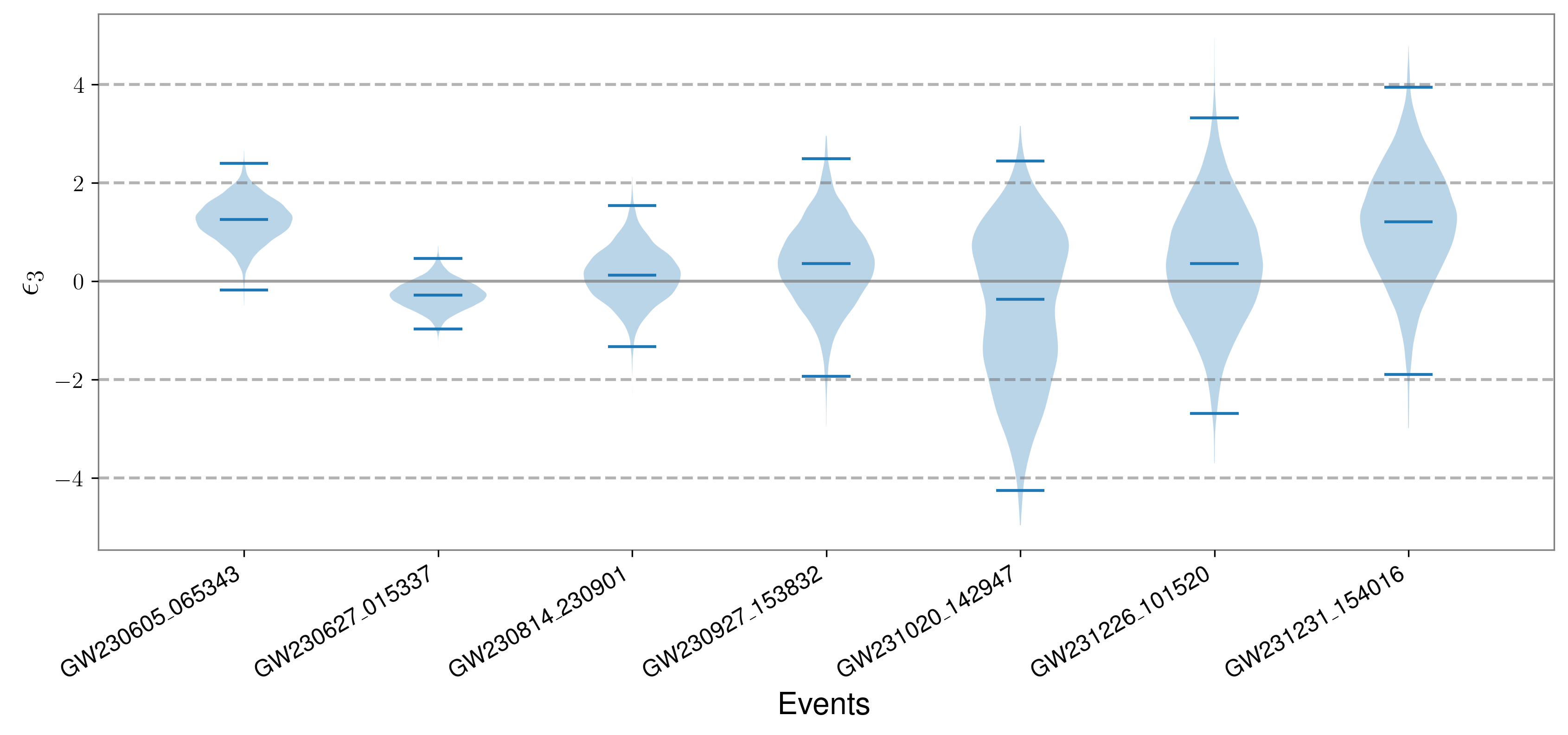}
\caption{Violin plots of constraints on $\alpha_{13}$ assuming $\epsilon_3 = 0$ (top panel) and $\epsilon_3$ assuming $\alpha_{13}= 0$ (bottom panel) from BBH events in GWTC-4. The gray solid line at 0 represents the expectation for black holes described by the Kerr metric. The error bars indicate the 3-$\sigma$ constraints of the non-Kerr parameters. \label{fig:def}} 
\end{figure*}

Overall, the GWTC-4 data significantly improve the precision of the constraints and do not qualitatively alter the conclusions regarding the Kerr hypothesis. The summary of the constraints is displayed in Fig.~\ref{fig:def}. Tab.~\ref{tab:99} shows the numerical values of the median of these posterior distributions along with the 99\% intervals. We note that there are four events (GW230628\_231200, GW230904\_051013, GW231108\_125142, and GW240109\_050431) that can provide only weak constraints on $\epsilon_3$ (their 99\% intervals exceed our range of $\epsilon_3$) and therefore these results are omitted in Fig.~\ref{fig:def} and Tab.~\ref{tab:99}.

\section{Concluding Remarks}\label{sec4}

In this work, we have extended our previous analysis of non-Kerr deviations in BBH inspirals to include data from the GWTC-4 catalog. Using the same theoretical framework and similar waveform modeling, we have assessed the impact of the expanded dataset on constraints of deformation parameters in a parametrized metric approach. Our results show that including GWTC-4 events yields tighter constraints on individual deformation parameters. The results remain consistent with the predictions of GR, with no statistically significant evidence of deviations from the Kerr metric.

\begin{figure*}[t]
\centering
\includegraphics[width =0.7\linewidth,trim=2.3cm 1.8cm 7.0cm 17.0cm,clip]{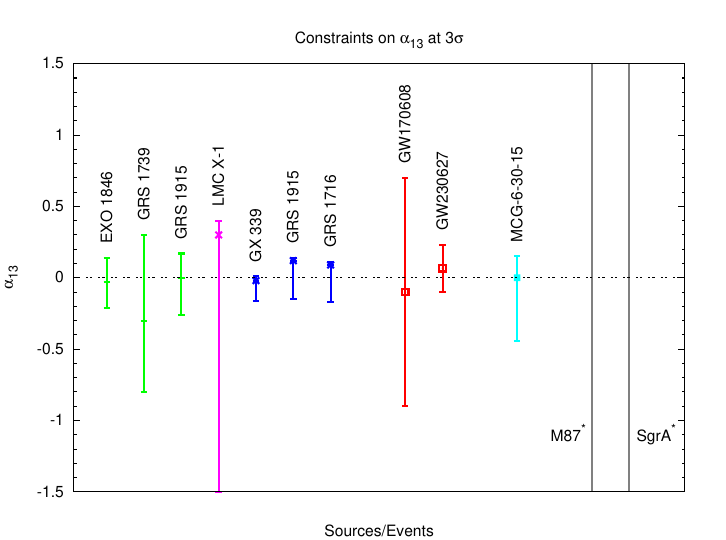}
\caption{Summary of the 3-$\sigma$ constraints on $\alpha_{13}$ with different techniques. Constrains from stellar-mass black holes: X-ray reflection spectroscopy (green), continuum-fitting method (magenta), X-ray reflection spectroscopy and continuum-fitting method (blue), and gravitational waves (red). Constrains from supermassive black holes:  X-ray reflection spectroscopy (cyan) and black hole imaging (gray). The label near every error bar indicates the name of the source/event used to infer the constraint. See the text for the details. \label{fig:summary}} 
\end{figure*}

Fig.~\ref{fig:summary} compares new and old GW constraints on $\alpha_{13}$ with electromagnetic constraints~\cite{2017RvMP...89b5001B} and updates Fig.~5.13 in Ref.~\cite{cb24}. Every error bar shows the 3-$\sigma$ constraint inferred from the source/event indicated by the label. We use different colors for the constraints inferred with different techniques. The error bars in green show the most precise and accurate measurements of $\alpha_{13}$ from stellar-mass black holes with X-ray reflection spectroscopy~\cite{2021ApJ...913...79T,2019ApJ...884..147Z}. The error bar in magenta is the measurements of $\alpha_{13}$ from the stellar-mass black hole in LMC~X-1 with the continuum-fitting method~\cite{2020ApJ...897...84T}. The error bars in blue are the measurements of $\alpha_{13}$ from the stellar-mass black holes in GX~339, GRS~1915, and GRS~1716 obtained by combining X-ray reflection spectroscopy and the continuum-fitting method~\cite{2021ApJ...907...31T,2022JCAP...01..019T,2022ApJ...924...72Z}. The error bars in red are the constraints from GWs: GW170608 is the event providing the most stringent constraint on $\alpha_{13}$ in GWTC-3~\cite{shashank2022constraining} and GW230627 is the event providing the most stringent constraint on $\alpha_{13}$ in GWTC-4 (and derived in the present work). The error bar in cyan shows the most precise and accurate measurement of $\alpha_{13}$ from supermassive black holes with X-ray reflection spectroscopy~\cite{2019ApJ...875...56T}. The error bars in gray, which exceed the range of $\alpha_{13}$ in our plot, are the constraints on $\alpha_{13}$ of M87$^*$ and SgrA$^*$ from the Event Horizon Telescope~\cite{2020PhRvL.125n1104P,2022ApJ...930L..17E}. While X-ray tests were able to provide the most stringent constraints on $\alpha_{13}$ before GWTC-4, GW tests have now reached the same constraining power and we can expect they will provide more stringent constraints on $\alpha_{13}$ than X-ray data in the next observing runs.

Our analysis strengthens the evidence that astrophysical black holes observed through GWs are consistent with the Kerr solution. Future GW observations, including the second and third parts of the fourth observing run, O4b and O4c~\cite{di2025status,bersanetti2026interferometer}, upcoming observing runs~\cite{abbott2020prospects} and next-generation detectors~\cite{danzmann2003lisa,luo2016tianqin,ziren2020introduction,Punturo:2010zz,Reitze:2019iox}, are expected to further improve the sensitivity to deviations from GR. Combining multiple events under the assumption of universal deformation parameters, or exploring alternative parametrizations, may provide a pathway to stronger constraints. The challenges and opportunities for future tests of gravity in the strong-field regime remain deeply intertwined, as each improvement in observational precision sharpens both our ability to confirm GR and our sensitivity to even the most subtle imprints of new physics.

\section*{Acknowledgements}

This work was supported by the National Natural Science Foundation of China (NSFC), Grant No.~W2531002. D.D. also acknowledges support from the China Scholarship Council (CSC), Grant No.~2022GXZ005434.

This research has made use of data or software obtained from the Gravitational Wave Open Science Center (gwosc.org), a service of the LIGO Scientific Collaboration, the Virgo Collaboration, and KAGRA. This material is based upon work supported by NSF's LIGO Laboratory which is a major facility fully funded by the National Science Foundation, as well as the Science and Technology Facilities Council (STFC) of the United Kingdom, the Max-Planck-Society (MPS), and the State of Niedersachsen/Germany for support of the construction of Advanced LIGO and construction and operation of the GEO600 detector. Additional support for Advanced LIGO was provided by the Australian Research Council. Virgo is funded, through the European Gravitational Observatory (EGO), by the French Centre National de Recherche Scientifique (CNRS), the Italian Istituto Nazionale di Fisica Nucleare (INFN) and the Dutch Nikhef, with contributions by institutions from Belgium, Germany, Greece, Hungary, Ireland, Japan, Monaco, Poland, Portugal, Spain. KAGRA is supported by Ministry of Education, Culture, Sports, Science and Technology (MEXT), Japan Society for the Promotion of Science (JSPS) in Japan; National Research Foundation (NRF) and Ministry of Science and ICT (MSIT) in Korea; Academia Sinica (AS) and National Science and Technology Council (NSTC) in Taiwan.

%\References
% references section
%\bibliographystyle{iopart-num}
\bibliography{ref}

\end{document}